\documentclass[]{aa}  
\usepackage{graphics}
\usepackage{psfig}
\begin{document}
   \thesaurus{08         
              (03.20.2;  
               08.09.2;  
               08.16.5;  
               08.02.1)  
             }
\title{Bispectrum speckle interferometry of the Orion Trapezium stars:
detection of a close (33 mas) companion of $\Theta^1{\rm Ori}$\,C\,\thanks{Based
on data collected at the SAO 6~m telescope in Russia.}}

 \author{Gerd Weigelt \inst{1}, Yuri Balega \inst{2}, Thomas Preibisch \inst{1},
 Dieter Schertl \inst{1}, Markus Sch\"oller \inst{3}, Hans Zinnecker \inst{4}} 

\offprints{weigelt@mpifr-bonn.mpg.de}

 \institute{Max-Planck-Institut f\"ur Radioastronomie,
 Auf dem H\"ugel 69, D--95121 Bonn, Germany \and
  Special Astrophysical Observatory, Nizhnij Arkhyz, Zelenchuk region, 
 Karachai-Cherkesia, 357147, Russia   \and
  European Southern Observatory, Karl-Schwarzschild-Str.~2, D--85748 Garching, Germany  \and
 Astrophysikalisches Institut Potsdam, An der Sternwarte 16,
D--14482 Potsdam, Germany}

   \date{Received 28 April 1999/ Accepted 5 June 1999}

   \maketitle
\markboth{Weigelt et al.:}{Bispectrum Speckle Interferometry of the Orion Trapezium}

   \begin{abstract}
We present bispectrum speckle interferometry observations 
with the SAO 6~m telescope of the four 
brightest stars in the Orion Trapezium. 
Diffraction-limited images with an unprecedented resolution 
$\lambda/D$ of \mbox{57~mas} 
and 76~mas were obtained in the $H$- and $K$-band, respectively. The $H$ and $K$
images of $\Theta^1{\rm Ori}$\,C  (the star responsible for the proplyds) 
show for the first time that $\Theta^1{\rm Ori}$\,C is a close binary with a 
separation of only $\sim\!33$ mas ($H$-band observation). The sub-arcsecond companions of $\Theta^1{\rm Ori}$\,A 
and $\Theta^1{\rm Ori}$\,B reported by Petr et al.~(1998) are confirmed. We use 
the magnitudes and colors of the companions to derive information about 
their stellar properties from the HR-diagram. In addition we briefly discuss 
the multiplicity of the Trapezium stars. Considering both, the visual and 
the spectroscopic companions of the 4 Trapezium stars, there are at least 7 
companions, i.e.~at least~1.75 companions per primary on average. 
This number is clearly higher than that
found for the low-mass stars in the Orion Nebula cluster as well as in
the field population. This suggests that a different mechanism is at 
work in the formation of high-mass multiple systems in the dense Trapezium 
cluster than for low-mass stars.

       \keywords{speckle interferometry; stars:~individual: $\Theta^1{\rm Ori}$\,C, 
 $\Theta^1 {\rm Ori}$\,A, $\Theta^1 {\rm Ori}$\,B;
 stars: binaries, stars: pre-main sequence}
   \end{abstract}


\section{Introduction}

The Orion Nebula cluster is one of the most prominent and nearby
($D \sim 450$ pc) star forming regions (for a review see Genzel \& 
Stutzki 1989). Its core contains a very dense cluster of young 
($\la 1\times 10^6$ yr) stars (cf.~Herbig \& Terndrup 1986;
McCaughrean \& Stauffer 1994; Hillenbrand 1997).
The Trapezium ($\Theta^1{\rm Ori}$\,ABCD), the system of the four most massive 
and luminous O-type and early B-type stars, is located in the center of 
the cluster. The strong stellar wind and the ionizing radiation of 
$\Theta^1{\rm Ori}$\,C has strong effects on the surrounding
cloud material (Bally et al.~1998; see also Richling \& Yorke 1998).

Petr et al.~(1998; P98 hereafter) presented the results of
130 mas resolution near-infrared speckle holographic observations
of the Trapezium cluster core, in which they could detect sub-arcsecond
 companions of the two Trapezium stars $\Theta^1{\rm Ori}$\,A and 
$\Theta^1{\rm Ori}$\,B. 
Simon et al.~(1999) 
reported the detection of an additional,
very faint companion of $\Theta^1{\rm Ori}$\,B.
In this paper we present the first near-infrared bispectrum 
speckle interferometry observations with diffraction-limited resolution of 
57 mas in the $H$-band and 76 mas in the $K$-band.

\section{Speckle observations and results}
 
\begin{figure}[h]
\parbox{8.8cm}{
\parbox{4.25cm}{\psfig{figure=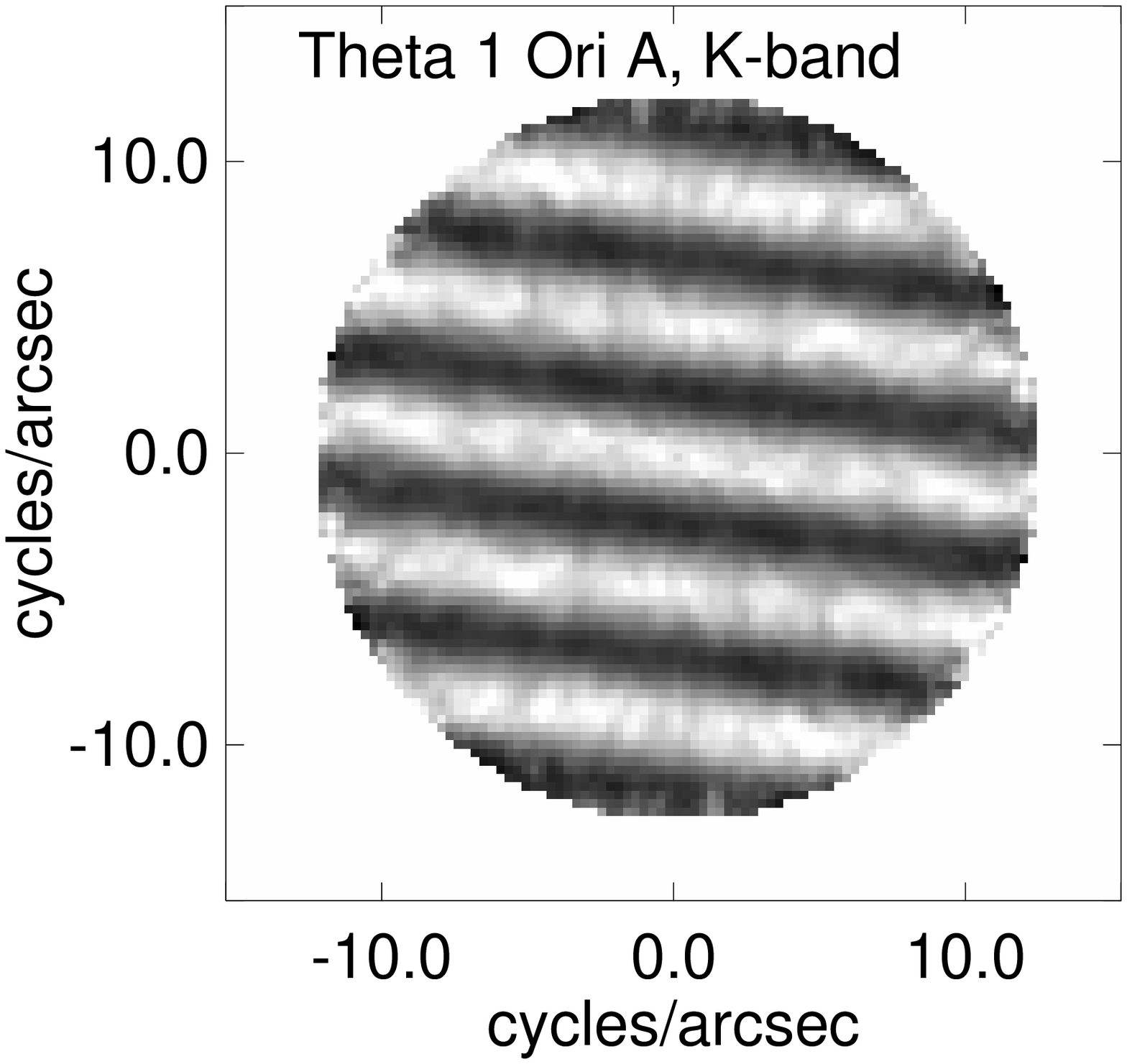,height=4.0cm,width=4.25cm}}\hfill
\parbox{4.25cm}{\psfig{figure=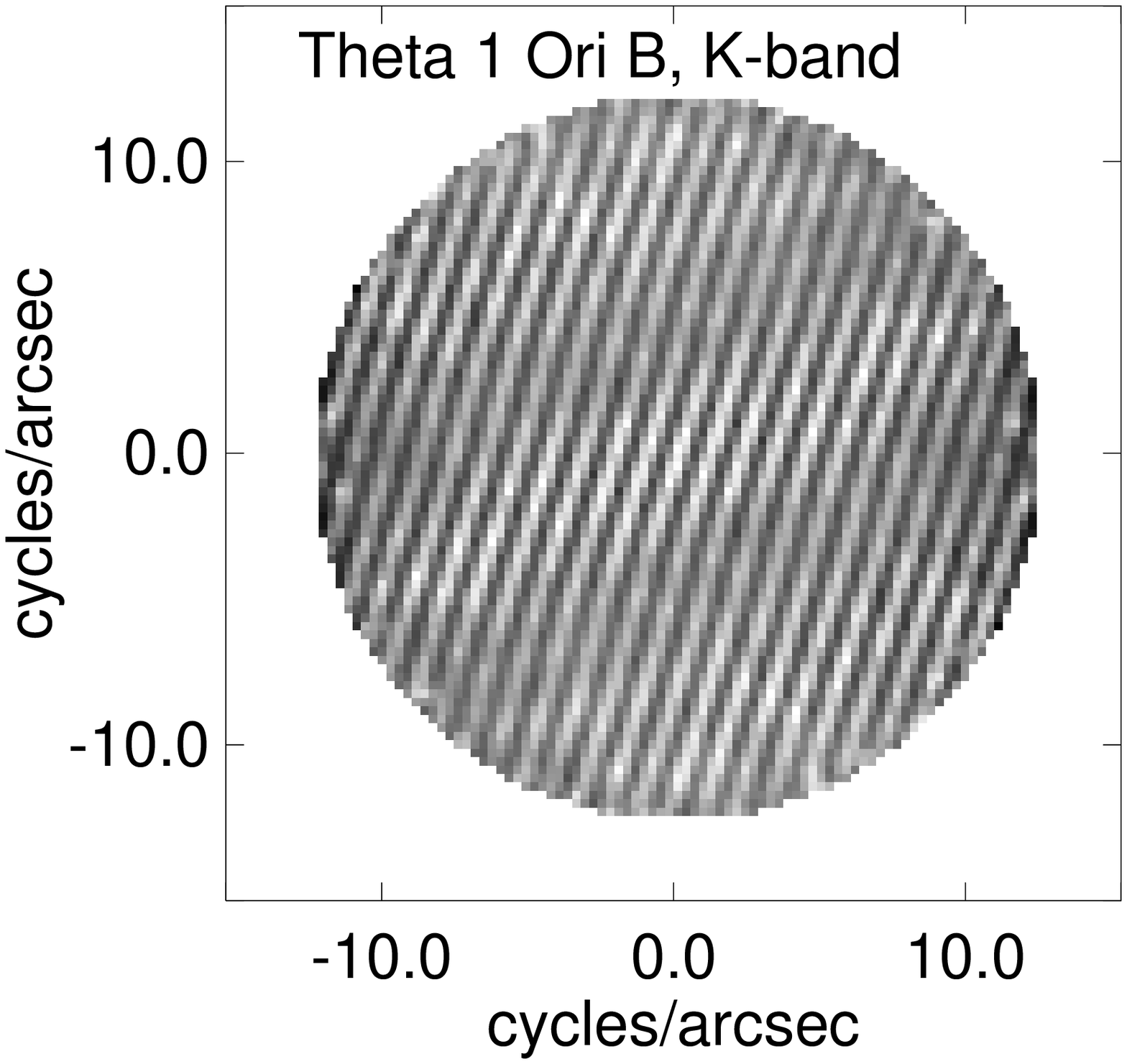,height=4.0cm,width=4.25cm}}}\vspace{1mm}

\parbox{8.8cm}{
\parbox{4.25cm}{\psfig{figure=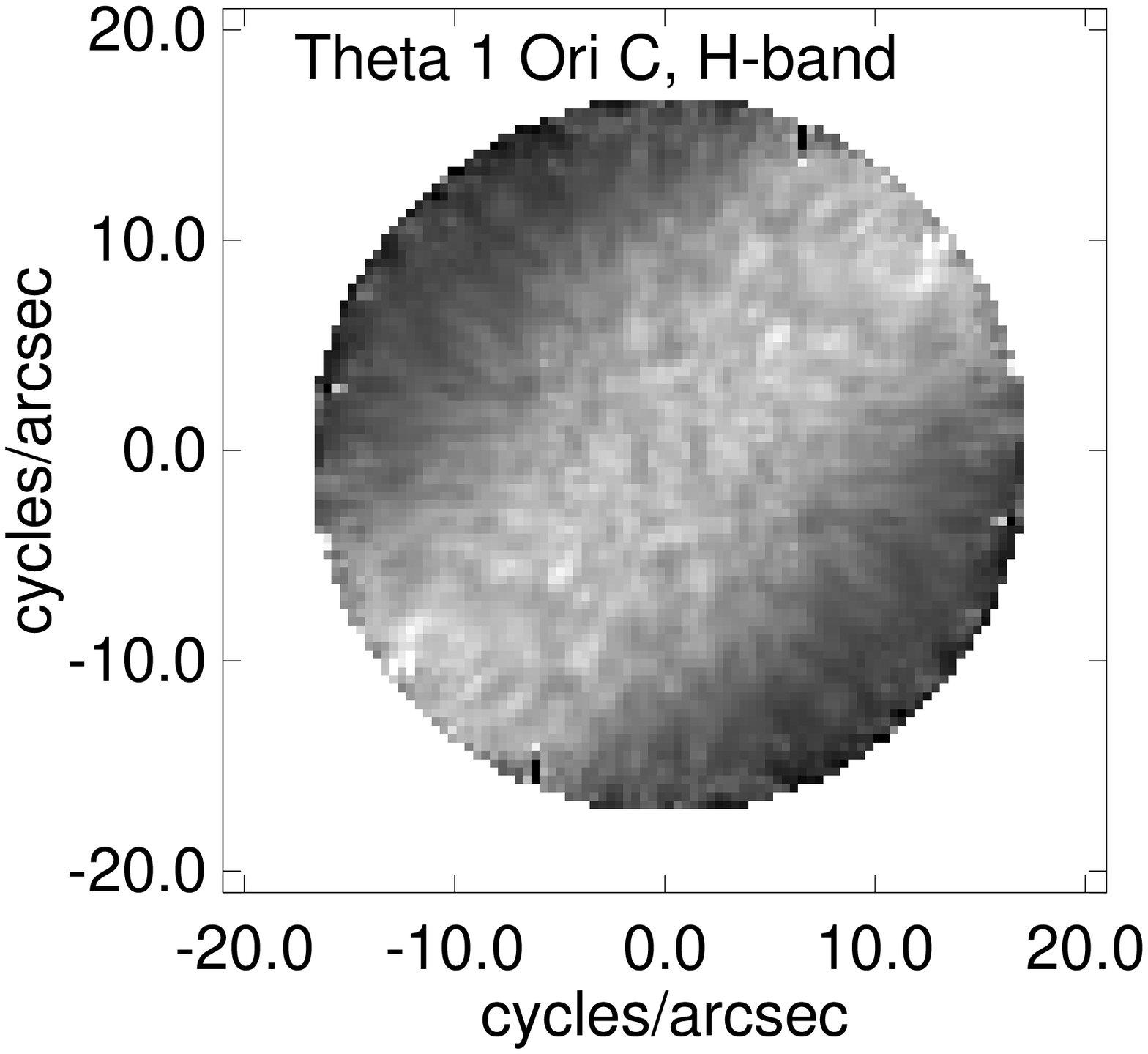,height=4.0cm,width=4.25cm}}\hfill
\parbox{4.25cm}{\psfig{figure=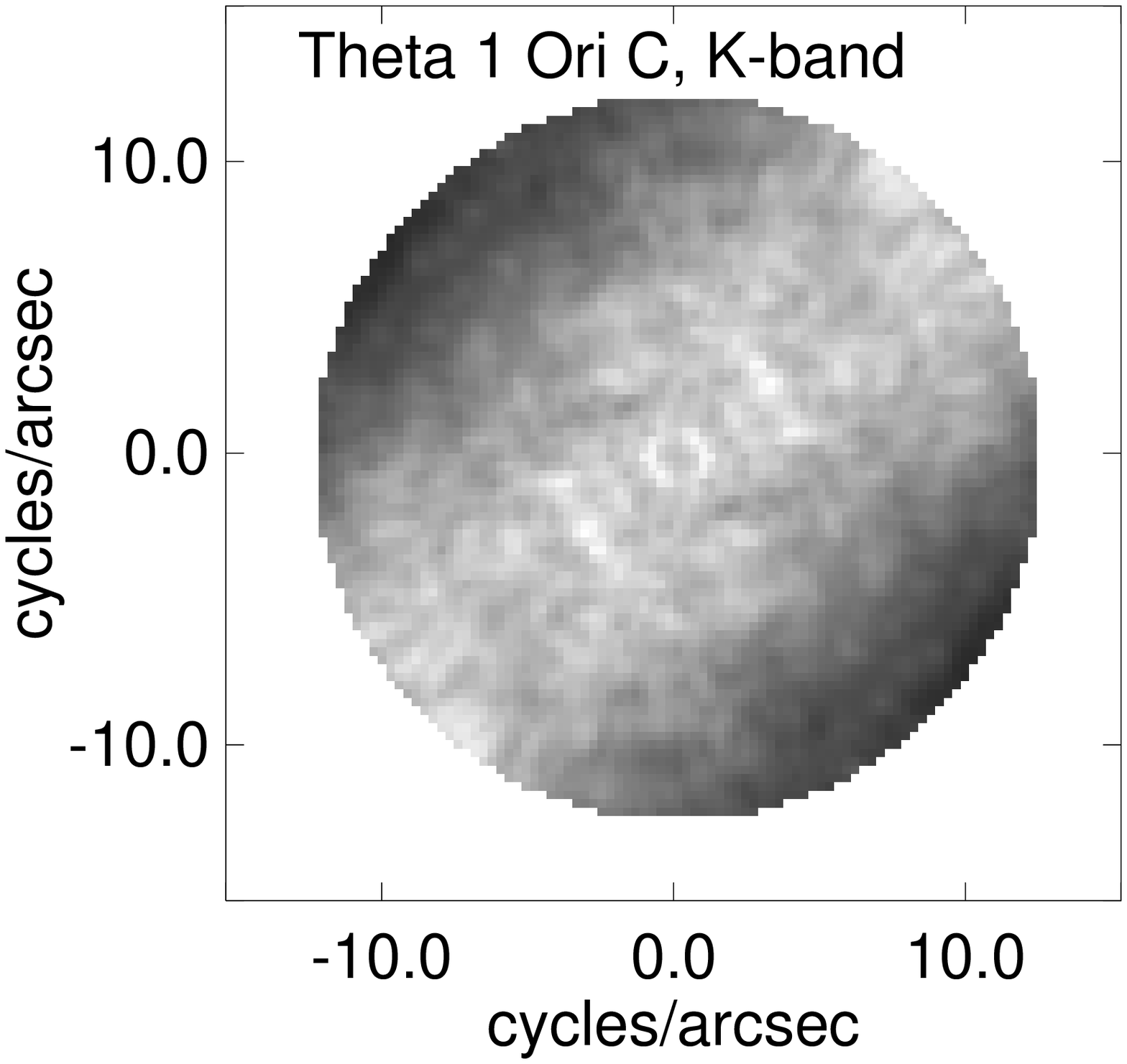,height=4.0cm,width=4.25cm}}}\vspace{1mm}

\parbox{8.8cm}{
\parbox{4.25cm}{\psfig{figure=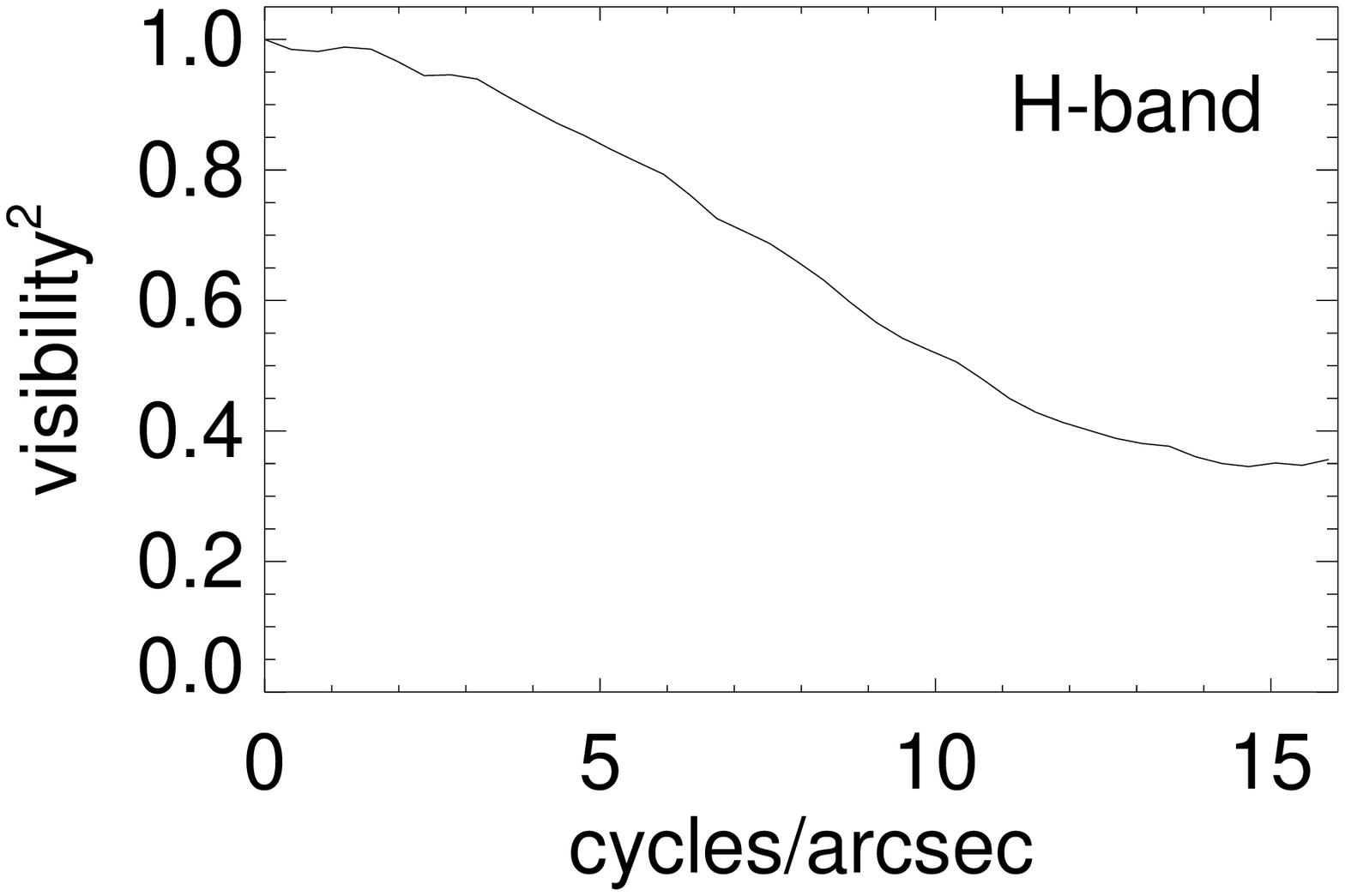,height=3.0cm,width=4.25cm}}\hfill
\parbox{4.25cm}{\psfig{figure=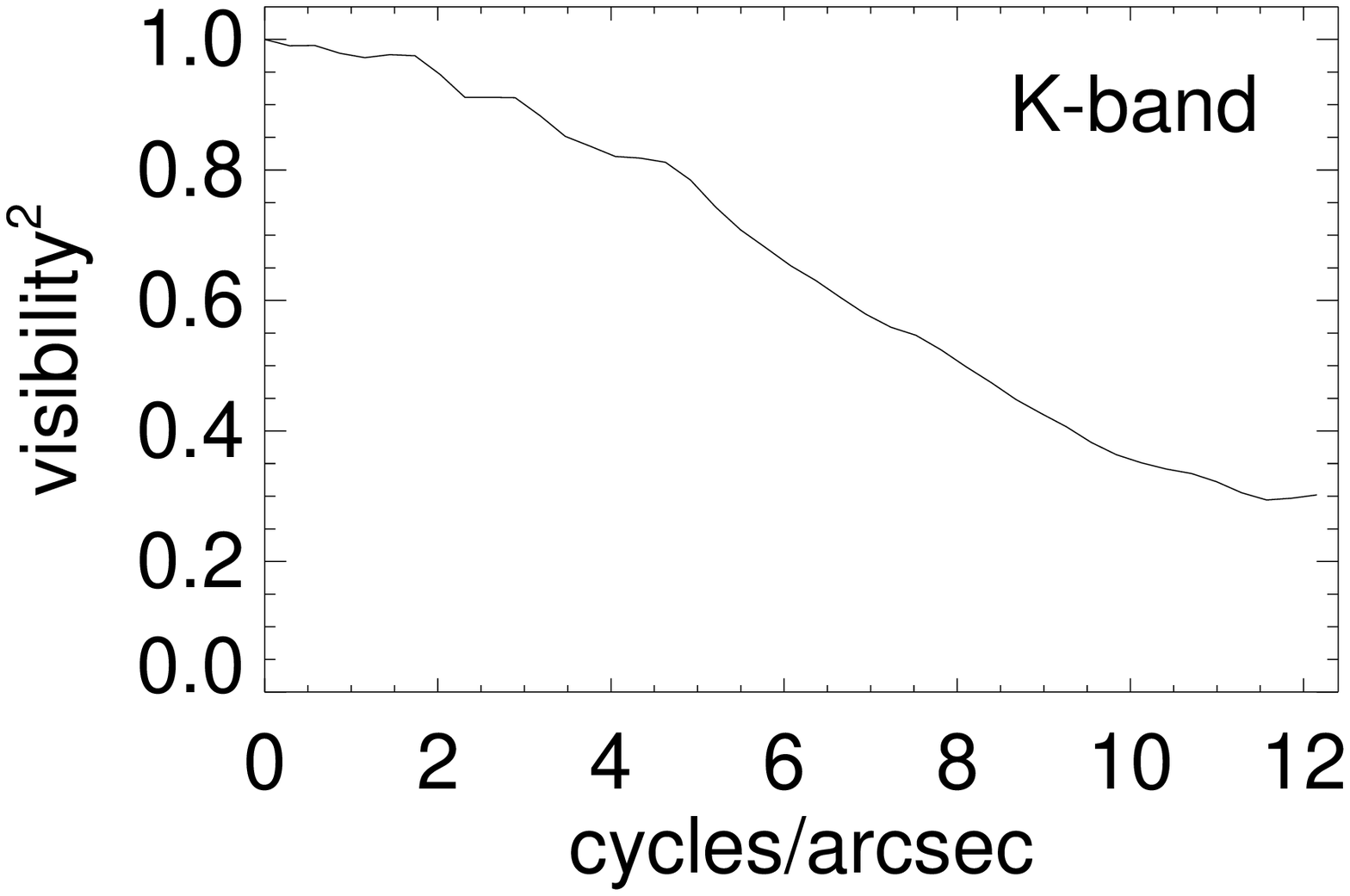,height=3.0cm,width=4.25cm}}}
\caption{Top: Reconstructed power spectra of $\Theta^1{\rm Ori}$\,A, B, and C;
Bottom: Line plots of the power spectra of $\Theta^1{\rm Ori}$\,C
perpendicular to the fringe direction.}
\label{ps}
\end{figure}

\begin{figure*} \centerline{
\parbox{17.84cm}{
\parbox{8.72cm}{\psfig{figure=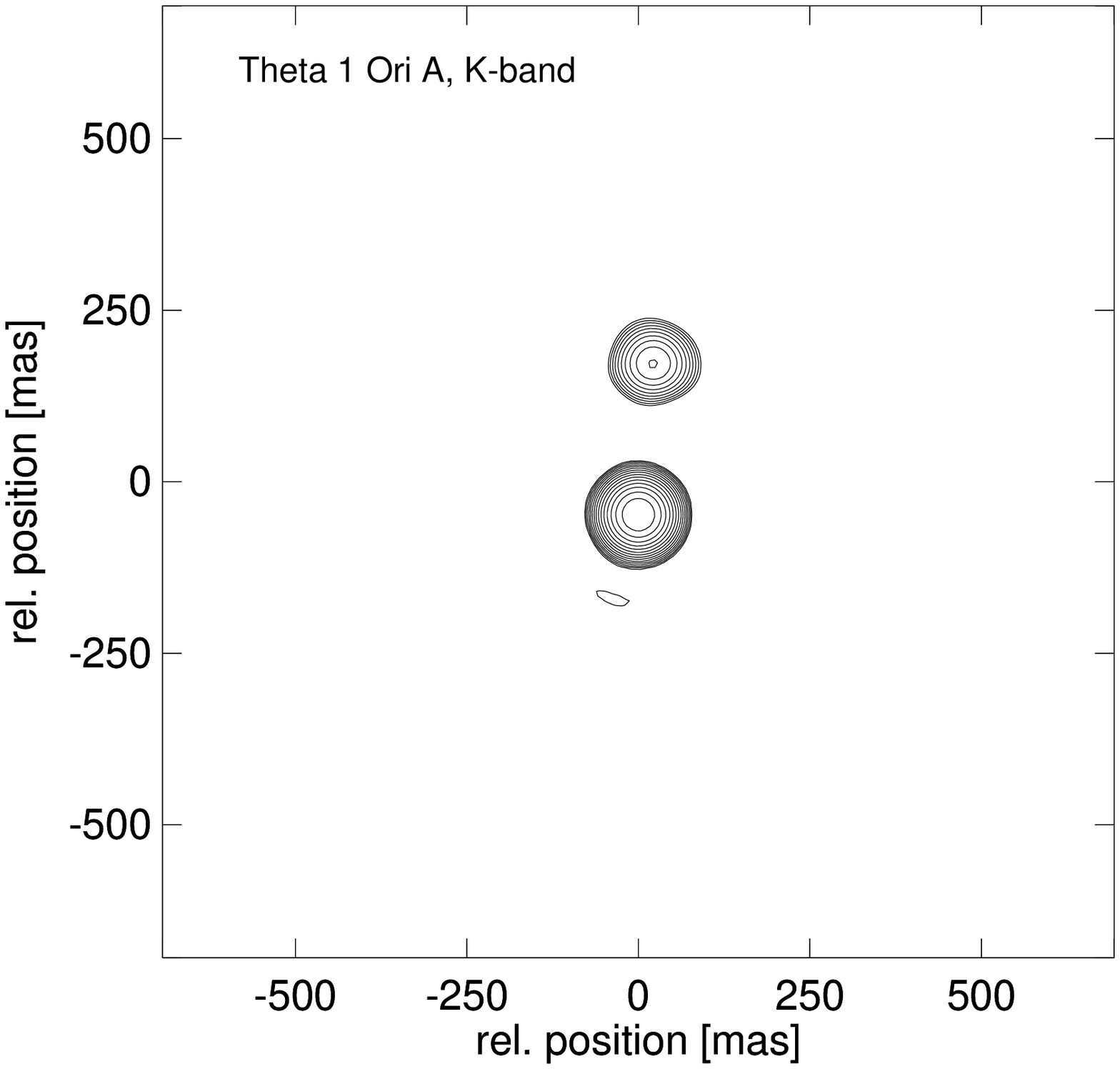,height=8.30cm,width=8.715cm}}\hspace{4mm}
\parbox{8.72cm}{\psfig{figure=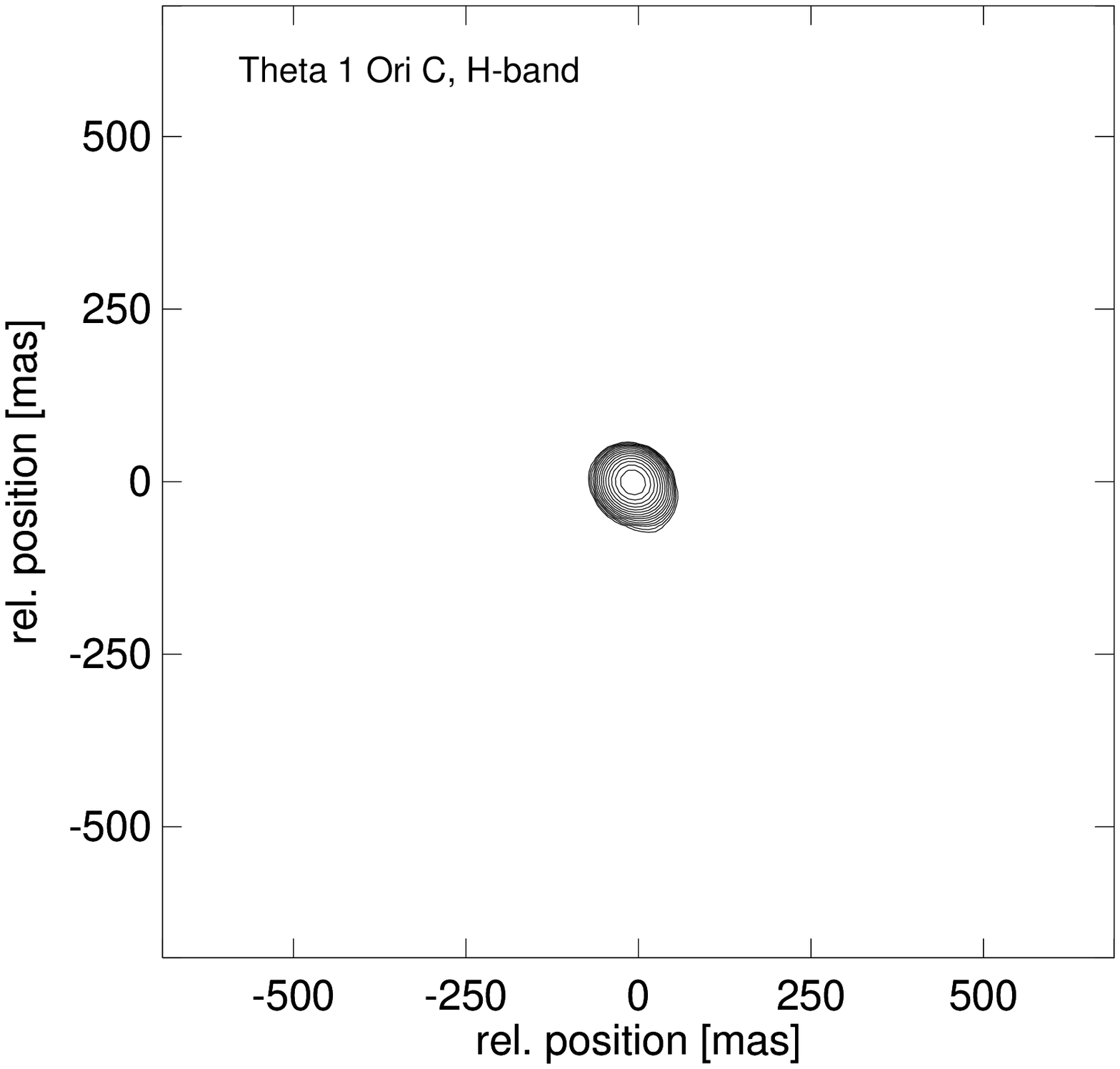,height=8.30cm,width=8.715cm}}}}
\vspace{1mm}

\centerline{
\parbox[b]{17.84cm}{
\parbox{8.72cm}{\psfig{figure=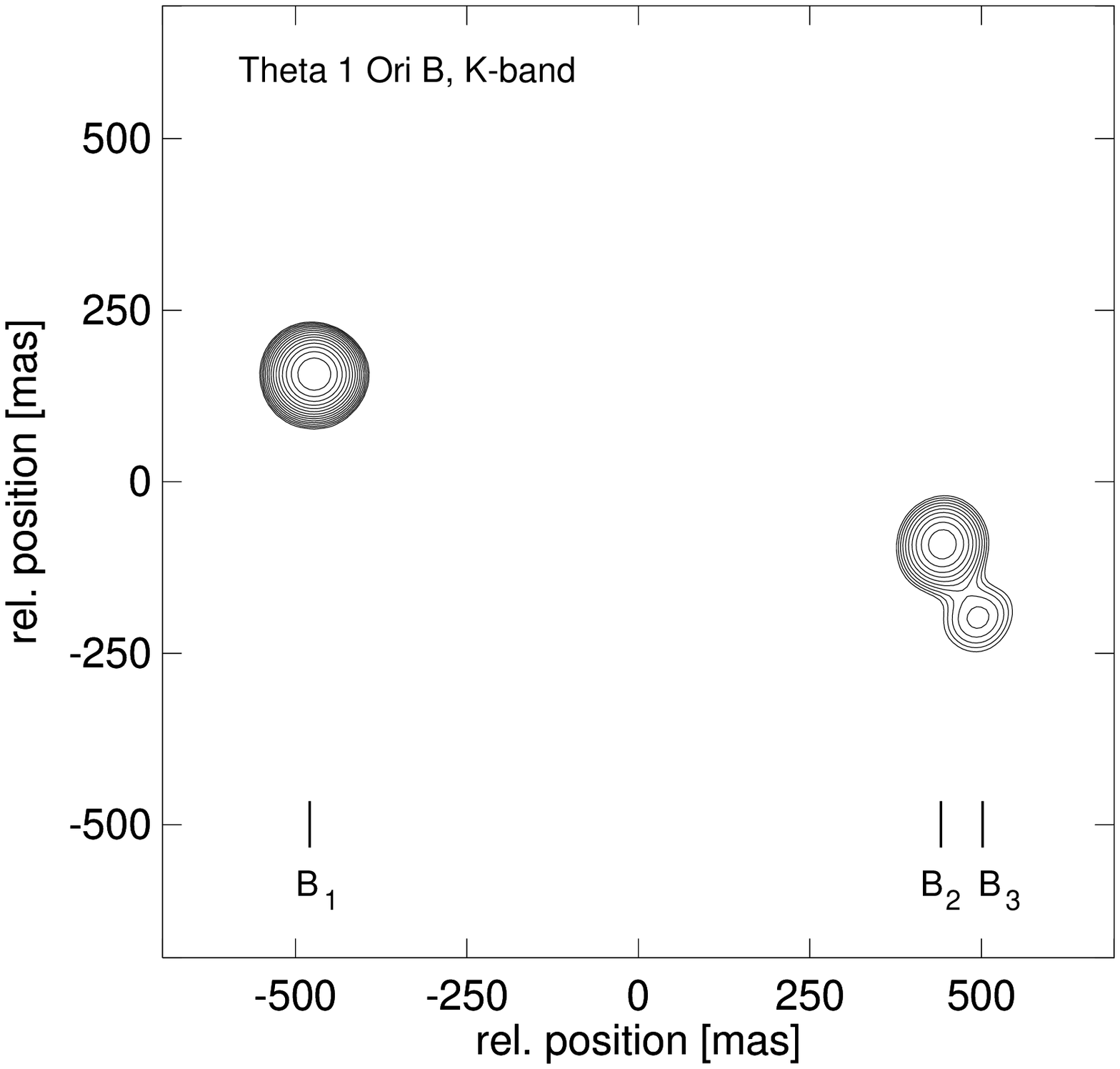,height=8.30cm,width=8.715cm}}\hspace{4mm}
\parbox{8.72cm}{\psfig{figure=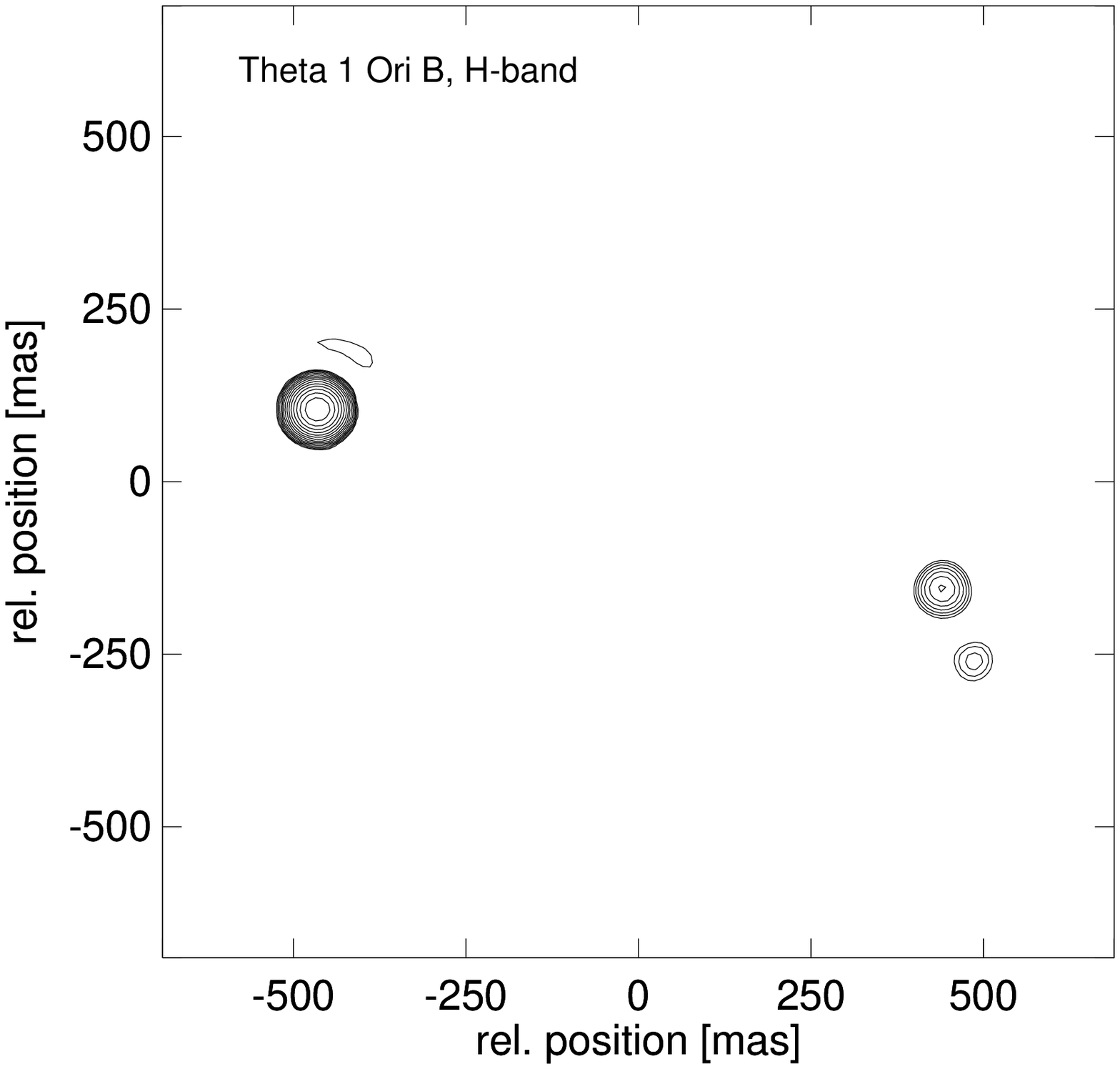,height=8.30cm,width=8.715cm}}}}
\caption{
Diffraction-limited images of $\Theta^1{\rm Ori}$\,A ($K$-band), B ($H$- and $K$-band), 
and C ($H$-band) reconstructed by the bispectrum speckle interferometry method. 
The contour level intervals are 0.25 mag, down to 3.75\,mag\,difference relative 
to the peak intensity. North is up and east is to the left.}
\label{images}
\end{figure*}

The 
speckle interferograms of $\Theta^1{\rm Ori}$\,A,~B,~C, and D 
were obtained with the 6~m telescope at the Special Astrophysical 
Observatory (SAO) in Russia on Oct.~14, 1997 ($H$-band) and 
Nov.~3, 1998 ($K$-band). Diffraction-limited images were 
reconstructed from the speckle data using the bispectrum speckle 
interferometry method (Weigelt 1977; Lohmann et al.~1983). The modulus of 
the object Fourier transform was determined with the speckle interferometry 
method (Labeyrie 1970). The speckle transfer functions were derived from 
speckle interferograms of the unresolved star $\Theta^1{\rm Ori}$\,E. Figures 
\ref{ps} and \ref{images}  show the power spectra and images 
of $\Theta^1{\rm Ori}$\,A, B, and C. 
The observational parameters and  the properties of the resolved stars 
are summarized in Table \ref{speckle_results}. 
The flux ratios  were determined by
fitting cosine functions to the power spectra.
\smallskip

\noindent
{\em $\Theta^1 Ori$\,A:}
The companion A$_2$ of the primary star $\Theta^1{\rm Ori}$\,A$_1$ 
(detected by P98) is clearly
visible in our images.\\
{\em $\Theta^1 Ori$\,B:}
The two companions $\Theta^1{\rm Ori}$\,B$_{2,3}$ are
clearly resolved, confirming the detection by P98.
In our images we cannot see the new faint 
component detected by Simon
et al.~(1999) because it is just below our detection limit.\\
{\em $\Theta^1 Ori$\,C:}
Our power spectra and images of $\Theta^1{\rm Ori}$\,C show 
a companion C$_2$ with a separation of $(33 \pm 5)$ mas from
the primary C$_1$ 
($H$-band; see Tab.~1).
 This is the
first detection of a companion of $\Theta^1{\rm Ori}$\,C.\\
{\em $\Theta^1 Ori$\,D:}
We find no indication for a companion of D. 

From the average surface density of stars in the Trapezium cluster reported
by Simon et al.~(1999) we estimate that the probability of finding a star
with $K < 10$ 
within $1''$ from a given position 
is $< 1\%$. This suggests
that the visual companions we observe actually are companions of
the respective primaries and  not only chance projections of unrelated stars.
\begin{table}
\caption[]{Observational parameters and results:\\
$H$-band:\,Nicmos3\,camera,\,$\lambda_c/\Delta \lambda$\,=\,1613/304\,nm, 
$600 \times 150$\,ms-~exposures, seeing $\sim 1.6''$ ,
scale 19.70\,mas/pixel;\\
$K$-band: Hawaii-array camera,
$\lambda_c/\Delta \lambda = 2160/320$ nm for A and B,
2200/200 nm for C,
$600 \times 120$ ms-exposures,
seeing $\sim 1.8''$, scale 27.00\,mas/pixel. \vspace{-1mm}

}
\begin{flushleft}
\begin{tabular}{l|ccc} 
$\Theta^1$ Ori          & Sep. [mas]   & Pos. Angle [deg.]  & Flux ratio     \\ \hline\hline
& \multicolumn{3}{c}{\it $H$-band observations (epoch 14 Oct. 1997):\rule[-1mm]{0mm}{4.5mm}}\\
$B_1B_2$ &$  942 \pm 20$      &$ 254.9 \pm 1$        &$ 0.12 \pm 0.02 $\\ 
$B_1B_3$ &$ 1018 \pm 20$      &$ 250.0 \pm 1$        &$ 0.06 \pm 0.03 $\\ 
$B_2B_3$ &$  114 \pm 5$       &$ 204.3 \pm 4$        &$ 0.40 \pm 0.04 $\\ 
$C_1C_2$ &$   33 \pm 5$       &$ 226   \pm 6$        &$ 0.26 \pm 0.02 $\\ \hline
&\multicolumn{3}{c}{\it $K$-band observations (epoch 03 Nov. 1998):\rule[-1mm]{0mm}{4.5mm}}\\
               $A_1A_2$ &$\ 221 \pm 5 $ &$ 353.8 \pm 2$       &$ 0.25 \pm 0.01 $\\ 
               $B_1B_2$ &$  942 \pm 20$ &$ 254.4 \pm 1$       &$ 0.30 \pm 0.02 $\\ 
               $B_1B_3$ &$ 1023 \pm 20$ &$ 249.5 \pm 1$       &$ 0.10 \pm 0.03 $\\ 
               $B_2B_3$ &$  117 \pm 5$  &$ 205.7 \pm 4$       &$ 0.32 \pm 0.05 $\\ 
               $C_1C_2$ &$   37 \pm 6$  &$ 222   \pm 5$       &$ 0.32 \pm 0.03 $\\ 
\end{tabular}
\label{speckle_results}
\end{flushleft} \vspace{-0.5cm}

\end{table} 

\section{Stellar properties of the companions}

\begin{figure*}
\centerline{\psfig{figure=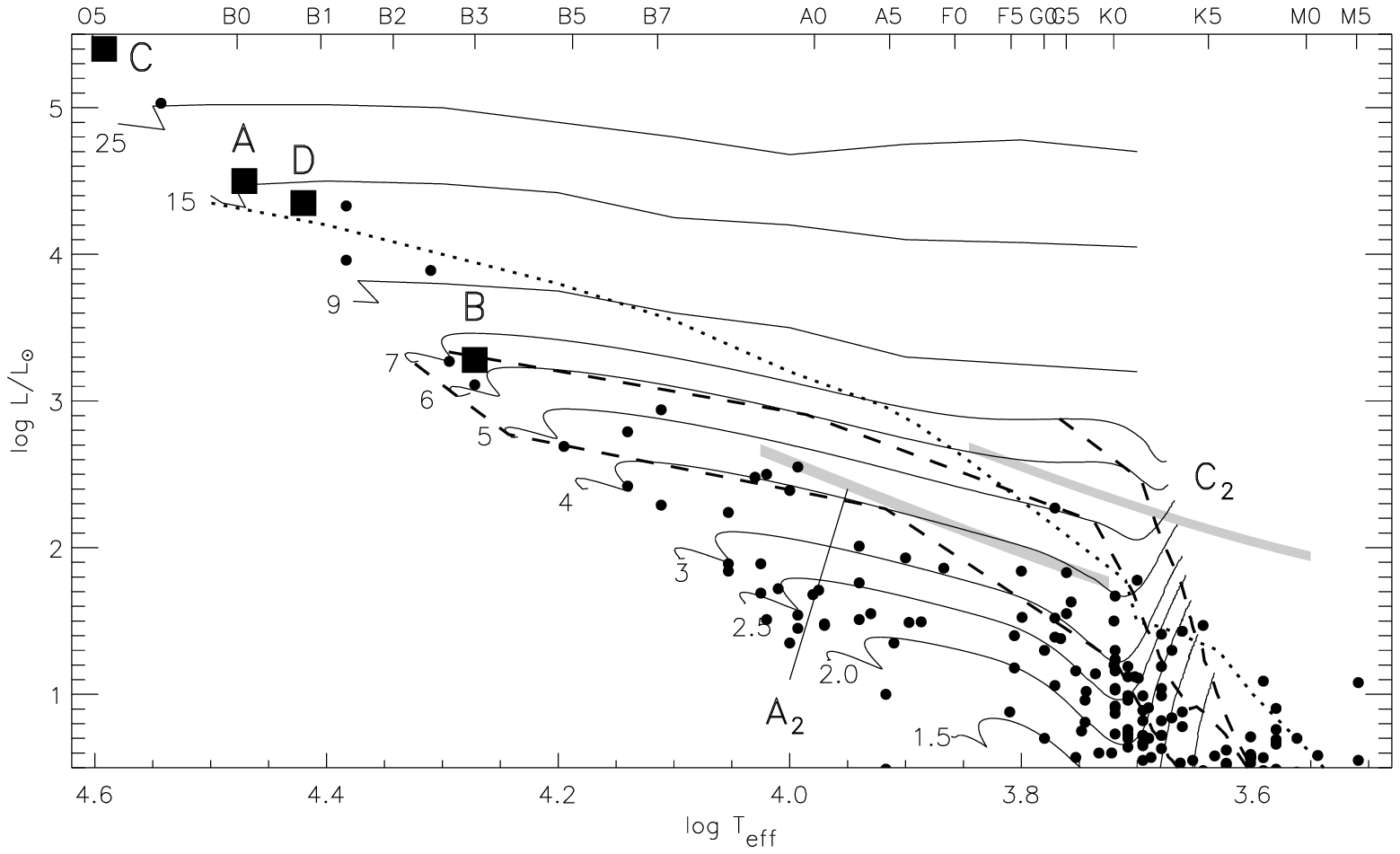,height=9.72cm,width=16.0cm}}
\caption{HRD with  PMS evolutionary tracks 
(labelled by the corresponding masses in $M_\odot$) from
Siess et al.~(1997) for $M \le 7\,M_\odot$ and from Bernasconi \& Maeder (1996)
for $M \ge 9\,M_\odot$.
The dashed lines show isochrones for ages 
of 0.1, 0.3, and 1 Myr. The dotted line shows the stellar birthline
(c.f.~Palla \& Stahler 1993)
for an accretion rate of $10^{-4}\,M_\odot\,{\rm yr}^{-1}$.
The positions of the primary stars are shown by the  
squares.  The grey bands show the locations of the speckle companions
A$_2$ and C$_2$, as described in the text.
The dots show other YSOs compiled from Hillenbrand (1997),
Preibisch (1999), and van den Ancker et al.~(1998).
}
\label{hrd}
\end{figure*}

In order to obtain information about the physical properties of the 
companions from our speckle results, we have 
 used the photometric and spectroscopic data and stellar parameters
compiled by Hillenbrand (1997) and Hillenbrand et al.~(1998).
From the known system magnitudes and the flux ratios determined in the
speckle images we have computed the $K$-band magnitudes and the $H-K$ colors 
of the speckle companions. These data can be used to estimate the luminosity
and the effective temperature of the stars: the $K$-band magnitude yields
the stellar luminosity as a function of the stellar temperature
(using the compilation of intrinsic $V-K$ colors and bolometric corrections
of Kenyon \& Hartmann 1995), and the $H-K$ color can be
transformed into the stellar temperature. The resulting
parameters can then be employed to estimate stellar masses and ages
(cf.~Fig.~\ref{hrd} and Table 2).

For $\Theta^1{\rm Ori}$\,C$_2$  we find
$K = 5.95 \pm 0.11$ and $H-K = 0.24\pm 0.10 $.
Since the extinction\footnote{We assume that the
extinction to the companion is the same as to the primary. Although one might
expect the companion, being a very young stellar object, to be surrounded by
 circumstellar 
material which might cause additional extinction, we believe that the strong 
radiation and wind of $\Theta^1{\rm Ori}$\,C would have dispersed any diffuse
material in its immediate vicinity very quickly.} 
 of $A_V = 1.8$ corresponds to $E(H-K) = 0.11$
(cf.~Rieke \& Lebofsky 1985),  
 the error ranges of the dereddened magnitude and color are 
$K_0 = [5.67 - 5.83]$ and $(H-K)_0 = [0.03\,-\,0.24]$. Since the magnitude
range defines a band in the HR-diagram 
and the color range corresponds
to a range of temperatures of $[3550 - 7000]$ K, these data 
define the grey shaded band in Fig.~\ref{hrd}.
The comparison of this band with theoretical
PMS tracks suggests that the companion  is a very young 
intermediate- or low-mass ($M \la 6\,M_\odot$) star.

For $\Theta^1{\rm Ori}$\,A$_2$ a similar computation (using the $H$-band flux
ratio from P98)  yields
$K_0 = [7.18 - 7.25]$ and $(H-K)_0 = [-0.01\,-\,0.09]$. The corresponding band
in the HRD suggests 
a mass of $M \approx 3\!-\!5\,M_\odot$.

The lack of photometric $H$-band data
for the components of $\Theta^1{\rm Ori}$\,B
prevents us from placing them into the HRD. Nevertheless, we can
derive upper limits for the masses from the $K$-band flux ratios 
if we assume that  the stars lie 
at or above, but not below, the main-sequence.
The corresponding limits are $M < 5\,M_\odot$ for B$_2$, 
$M < 3.5\,M_\odot$ for B$_3$, and 
$M < 2\,M_\odot$ for B$_4$ (the faint companion detected by Simon et
al.~1999).

\section{Multiplicity of the massive Trapezium stars}

Several recent studies (P98; Simon et al.~1999)
have concordantly found that the binary frequency of the
low-mass stars in the Orion nebula cluster (ONC) is
comparable to that of solar type field stars, which is 
about 60\% with a median number of companions per primary 
of about 0.5 (c.f.~Duquennoy \& Mayor 1991; Fischer \& Marcy 1992).
The binary frequency of O-type and early B-type field stars seems to be similar
(cf.~ Abt 1983; Mason et al.~1998).

Our detection of $\Theta^1{\rm Ori}$\,C$_2$
increases the number of known companions to the four
Trapezium stars to 7.
The average number of at least 1.75 companions per primary among the high-mass
Trapezium stars is clearly higher than the 
corresponding number for the low-mass stars in the
ONC. 
A similar trend has been found by  Abt et al.~(1991) and 
Morrell \& Levato (1991): most of the spectroscopic binaries in the ONC
are among the O- and early B-type stars, and much
less frequent among the later B- and A-type stars.
This finding suggests different formation mechanisms for the high-mass
and low-mass multiple systems.
This is consistent with the recent results of Bonnell et al.~(1998)
who assumed that high-mass stars form through  accretion-induced collisions
of protostars. Their theory predicts that close binary systems should be
very common among the massive stars. This is supported by our results.

\begin{table}[h]
\caption[]{Companions of the Trapezium stars ($M_p$: primary mass,
$M_s$: secondary mass, $\rho$: distance). References:
 1: this work; 2: P98; 3: Bossi et al.~(1989); 4: Simon et al.~(1999);
 5: Abt et al.~(1991)}
\begin{flushleft}
\begin{tabular}{cc|lccr}
Primary & $M_p$ & Comp. & $\rho$ & $M_s/M_p$ & Ref. \\
$\Theta^1$ Ori & $[M_\odot]$ & & [AU] & & \\ \hline
A$_1$ & 20 &A$_2$ (vis)& 100  & $\sim 0.2$ & 1,2 \\
  &    &A$_3$ (spec)&  1  & $\sim 0.13$ & 3 \\ \hline
B$_1$ &  7 &B$_2$ (vis)& 430  & $< 0.7$ & 1\\
  &    &B$_3$ (vis)& 460  & $< 0.5$ & 1\\
  &    &B$_4$ (vis)& 260  & $< 0.3$ & 4\\
  &    &B$_5$ (spec)&0.13 &   & 5\\ \hline
C$_1$ & 45 &C$_2$ (vis)&  16  & $< 0.13$ & 1\\ \hline
D & 17 &       \multicolumn{3}{l}{apparently single} \\
\end{tabular}
\label{companions}
\end{flushleft}
\end{table}

\end{document}